\documentclass[12pt]{article}
\usepackage{epsfig}
\usepackage{cite}

\begin{document}
\begin{titlepage}

\centerline{\large\bf Center vortex model for Sp(2) Yang-Mills theory}

\bigskip
\centerline{M.~Engelhardt and B.~Sperisen}
\vspace{0.2 true cm}
\centerline{\em Physics Department, New Mexico State University}
\centerline{\em Las Cruces, NM 88003, USA}

\abstract{The question whether the center vortex picture of the strongly
interacting vacuum can encompass the infrared dynamics of both $SU(2)$
as well as $Sp(2)$ Yang-Mills theory is addressed. These two theories
contain the same center vortex degrees of freedom, and yet exhibit
deconfinement phase transitions of different order. This is argued to be
caused by the effective action governing the vortices being different in
the two cases. To buttress this argument, a random vortex world-surface
model is constructed which reproduces available lattice data 
characterizing $Sp(2)$ Yang-Mills confinement properties. A new
effective action term which can be interpreted in terms of a vortex
stickiness serves to realize a first-order deconfinement phase transition,
as found in $Sp(2)$ Yang-Mills theory. Predictions are given for the
behavior of the spatial string tension at finite temperatures.}

\vspace{1cm}



\end{titlepage}

\section{Introduction}
The random vortex world-surface model describes the infrared,
nonperturbative regime of the strong interaction on the basis of effective
gluonic center vortex degrees of freedom. Such a description was initially
suggested and studied in \cite{hooft,aharonov,cornold,mack,olesen} in
particular with a view towards explaining the confinement phenomenon; more
recently, investigations of the relevance of center vortices in the lattice
Yang-Mills ensemble \cite{jg1,jg2,tk1,df1,int,per}, for a review,
cf.~\cite{jg3}, have provided a firm foundation for this picture. Motivated
by these results, random vortex world-surface models have been formulated
and studied both with respect to $SU(2)$ as well as $SU(3)$ Yang-Mills
theory \cite{m1,m2,m3,su3conf,su3bary,su3freee}, successfully reproducing
the main features of the strongly interacting vacuum. In the $SU(2)$ case,
not only has a confining low-temperature phase been obtained together
with a second-order deconfinement phase transition as temperature is
raised \cite{m1}; also the topological susceptibility
\cite{m2,cw2,contvort,rb,bruck} and the (quenched)
chiral condensate \cite{m3} of $SU(2)$ Yang-Mills theory are reproduced
quantitatively. In the $SU(3)$ case, the deconfinement transition becomes
weakly first order \cite{su3conf} and a Y-law for the baryonic static
potential results in the confining phase \cite{su3bary}.

Rather than immediately pursuing the next logical step in this development,
namely, extending the $SU(3)$ investigation to the topological and chiral
properties, recent efforts have focused on the question of how far the
simple random vortex world-surface concept carries if one generalizes
to other gauge groups. The most obvious extension, to the $SU(4)$ group,
was reported in \cite{su4}. This study indeed confirmed the expectation
formulated in \cite{jeffstef}: As the number of colors $N$ is increased,
Abelian magnetic monopoles, which are an intrinsic feature of generic
center vortices, begin to influence the distribution of vortex
configurations instead of being completely enslaved to the dynamics
of the vortices which host them. Unequivocal signatures of this emerge
in constructing the $SU(4)$ random vortex world-surface ensemble.
Physically, the reason for this behavior is rooted in the fact that
the flux emanating from Abelian magnetic monopoles is quantized in
identical units for any $N$, whereas the flux carried by center vortices
is quantized in ever smaller units as $N$ rises. The vortices thus
become ``lighter'' degrees of freedom in relation to the monopoles;
the latter then attain a dynamical significance of their
own\footnote{It should be emphasized that this does not imply that vortices
cease to represent the relevant infrared degrees of freedom as the
number of colors rises; all that happens is that their dynamics become
more complex, and cannot be described purely in terms of world-surface
characteristics anymore.}.

On the other hand, another systematic way of extending the Yang-Mills
gauge group has recently also garnered attention \cite{spn,g21o,pepelat05}:
The $SU(2)$ group can alternatively be viewed as the smallest symplectic
group $Sp(1)$, and the sequence of $Sp(N)$ groups can also be used as a
systematic generalization of $SU(2)=Sp(1)$. An interesting aspect of this
sequence is that all $Sp(N)$ have the same center, $Z(2)$; furthermore,
all gauge groups $Sp(N)$ have the same first homotopy group after
factoring out the center, $\Pi_{1} (Sp(N)/Z(2))=Z(2)$. This means
that they allow for the same set of center vortex degrees of freedom.
The studies \cite{spn,g21o} report lattice investigations of selected
$Sp(N)$ Yang-Mills theories; the data gathered there now provide an
opportunity to confront the random vortex world-surface model with
these theories. In particular, while $SU(2)=Sp(1)$ Yang-Mills theory
exhibits a second-order deconfinement phase transition, the transition
is first order in $Sp(2)$ Yang-Mills theory. Thus, one has two
Yang-Mills theories with the same center and center vortex content
which display completely different behavior at the deconfinement
transition. This raises the question whether center vortices indeed are
the relevant degrees of freedom determining the physics of confinement
and, in particular, the transition to a deconfined high-temperature
phase. Of course, whereas the infrared effective vortex models
corresponding to $SU(2)$ and $Sp(2)$ Yang-Mills theory are based on
the same set of vortex degrees of freedom, the respective effective
vortex actions may be quite different; after all, they formally result
from integrating out the very different cosets of the two gauge groups.
Thus, a different behavior of the two models at the deconfinement transition
is by no means excluded. Nevertheless, it would be useful to buttress
this argument by an explicit construction of a random vortex world-surface
model for $Sp(2)$ Yang-Mills theory, to demonstrate that the vortex
picture can encompass the confinement physics of both the $SU(2)$ and
the $Sp(2)$ cases. To furnish such a construction is the objective of
the present work.

\section{Sp(2) lattice Yang-Mills theory data}
\label{sp2data}
The objective of the present investigation is to find a $Z(2)$-symmetric
random vortex world-surface model with a first-order deconfinement
phase transition and, if possible, adjust it to reproduce known data
on $Sp(2)$ Yang-Mills theory. Two relevant quantitative characteristics
are reported in \cite{spn}, namely, the ratio of the deconfinement
temperature to the square root of the zero-temperature string tension,
$T_c /\sqrt{\sigma } $, and the latent heat $L_H $. The latent heat
corresponds to the discontinuity in the four-dimensional action
density\footnote{Since the symbol $s$ will serve a different purpose below,
the action density is denoted $\bar{s} $ here and in the following.}
$\bar{s} $ at the first-order deconfinement transition, and is given in
\cite{spn} in units of the lattice spacing $a$, i.e.,
$L_H = a^4 \Delta \bar{s} $. While \cite{spn} gives $T_c /\sqrt{\sigma } $
for a number of $Sp(2)$ lattice Yang-Mills couplings and the extrapolation
to the continuum limit, $L_H $ is only reported quantitatively for one
coupling, $8/g^2 = 6.4643$; it should be noted that the scaling regime does
not quite extend to that strong a coupling. In view of these data,
it seems most consistent to model $Sp(2)$ Yang-Mills theory specifically
at the aforementioned coupling, $8/g^2 = 6.4643$, as opposed to using a
mixed input data set consisting of the continuum limit of
$T_c /\sqrt{\sigma } $ on the one hand and the value of $L_H $
at $8/g^2 = 6.4643$ on the other hand.

At $8/g^2 = 6.4643$, one has \cite{spn}
\begin{equation}
T_c /\sqrt{\sigma } = 0.59
\label{input1}
\end{equation}
(in the continuum limit, this value rises to $0.69$). On the other hand,
identifying $L_H = a^4 \Delta \bar{s} $, the action density discontinuity
$\Delta \bar{s} $ satisfies \cite{spn}
\begin{equation}
N_t (a^4 \Delta \bar{s} )^2 /4 =0.15
\end{equation}
where $N_t $ denotes the extent of the lattice in the (Euclidean) time
direction. Taking into account that, at this coupling, the deconfinement
transition occurs at $N_t =2$, i.e., the deconfinement temperature
is given by $T_c a=1/2$, one can eliminate the lattice spacing,
yielding
\begin{equation}
\Delta \bar{s} / T_c^4 = 8.76
\label{input2}
\end{equation}
The two relations (\ref{input1}) and (\ref{input2}) will serve as input
data for the random vortex world-surface model constructed below.

\section{Random vortex world-surface model}
Center vortices are closed tubes of quantized chromomagnetic flux in
three spatial dimensions. In four-dimensional (Euclidean) space-time,
they are therefore represented by (thickened) world-surfaces. The
quantization of flux is defined by the center of the gauge group; a
Wilson loop linked to a vortex yields a nontrivial center element
(the trivial unit element signals absence of any flux). For gauge
groups with a $Z(2)$ center, such as the $SU(2)$ case studied in
\cite{m1,m2,m3} or the $Sp(2)$ case studied here, this implies that
there is only one type of vortex flux, corresponding to the only
nontrivial center element $(-1)$.

The model construction used in the following is entirely analogous to the
$SU(2)$ model \cite{m1}, and the reader is referred to that work for further
details regarding the construction and interpretation of random vortex
world-surface models. As argued in the introductory section further above,
differences between the $SU(2)$ and $Sp(2)$ models arise only at the level
of the effective vortex action, discussed further below. Apart from that
discussion, thus, a brief overview of the modeling methodology shall
suffice:

In order to arrive at a tractable model description, vortex world-surfaces
are composed of elementary squares on a hypercubic lattice. The lattice
square extending from the site $x$ into the positive $\mu $ and $\nu $
directions (where, for definiteness, $\mu < \nu $) is associated with a
value $q_{\mu \nu } (x) \in \{ 0,1 \} $, where the value $1$ means that
the square is part of a vortex surface and the value $0$ means it is
not\footnote{Note that this is adequate for the description of confinement
properties, since the Wilson loop is insensitive to flux orientation. To
treat topological and chiral properties, a slightly extended description,
which permits the specification of vortex orientation, is needed
\cite{m2,m3}.}. For ease of notation below, it is useful to define also
$q_{\nu \mu } (x) = q_{\mu \nu } (x) $. An ensemble of vortex world-surface
configurations is generated by Monte Carlo update. In order to preserve
the closed character of the world-surfaces, an elementary update acts on
all six squares forming the surface of an elementary three-dimensional
cube in the four-dimensional lattice. If the cube extends from the site
$x$ into the positive $\mu $, $\nu $ and $\lambda $ directions, then an
update simultaneously effects
\begin{equation}
\begin{array}{ll}
q_{\mu\nu}(x) \rightarrow 1-q_{\mu\nu}(x)
\, , \ \ \ \ \ \ &
q_{\mu\nu}(x+e_\lambda) \rightarrow 1-q_{\mu\nu}(x + e_\lambda) \, , \\
q_{\nu \lambda}(x) \rightarrow 1-q_{\nu \lambda}(x)
\, , \ \ \ \ \ \ &
q_{\nu \lambda}(x+e_\mu) \rightarrow 1-q_{\nu \lambda}(x+e_\mu) \, , \\
q_{\lambda\mu}(x) \rightarrow 1-q_{\lambda\mu}(x)
\, , \ \ \ \ \ \ &
q_{\lambda\mu}(x+e_\nu) \rightarrow 1-q_{\lambda\mu}(x+e_\nu) \, .
\end{array}
\end{equation}
In practice, sweeps through the lattice are performed in which updates
involving all three-dimensional cubes in the lattice are considered in
turn.

One important point which should be noted is the interpretation of the
lattice spacing \cite{m1}. In random vortex world-surface models, the
lattice spacing is a fixed physical quantity, which in the $SU(2)$ and
$SU(3)$ cases is determined \cite{m1,su3conf} to be $0.39\, $fm (where
the scale is set by defining the zero-temperature string tension $\sigma $
to be $\sigma = (440\, \mbox{MeV} )^2 $). Physically, this introduces into
the models the notion that vortices possess a certain transverse thickness.
While they are formally represented as two-dimensional surfaces, the
fixed lattice spacing prevents, e.g., two parallel vortices from propagating
at such a short distance from one another that they would cease to be
mutually distinguishable if their transverse profile were explicitly taken
into account. Random vortex world-surface models are thus infrared
effective theories with a fixed ultraviolet cutoff in form of a fixed
lattice spacing determined by the physical vortex thickness.

As mentioned above, the substantive difference between the $SU(2)$ vortex
model studied in \cite{m1,m2,m3} and the $Sp(2)$ vortex model arises at
the level of the vortex effective action used in the Monte Carlo
generation of the vortex ensemble. In the $SU(2)$ (and also the
$SU(3)$) model, one action term (and, therefore, one adjustable
dimensionless parameter) is sufficient to achieve quantitative agreement
between infrared observables studied in the vortex model and the data
from the corresponding full lattice Yang-Mills theory. The action term
in question is a world-surface curvature term,
\begin{eqnarray}
S_c [q] \! \! \! \! &=& \! \! \! \!
c \sum_x\sum_\mu \left[ \sum_{\nu < \lambda \atop \nu \neq \mu,
\lambda\neq \mu} \left( q_{\mu\nu}(x) \, q_{\mu\lambda}(x)
+ q_{\mu\nu}(x) \, q_{\mu\lambda}(x-e_\lambda)
\right. \right. \label{curvature} \\
& & \ \ \ \ \ \ \ \ \ \ \ \ \ \ \ \ \ \
+ \left. q_{\mu\nu}(x-e_\nu) \, q_{\mu\lambda}(x)
+ q_{\mu\nu}(x-e_\nu) \, q_{\mu\lambda}(x-e_\lambda)
\right)\Bigg] \nonumber \\
&=& \! \! \! \! \frac{c}{2} \sum_{x}\sum_\mu \left[ \left[
\sum_{\nu\neq\mu} \left( q_{\mu\nu}(x) +
q_{\mu\nu}(x-e_\nu) \right) \right]^2 \! \! \! - \! \!
\sum_{\nu\neq\mu} \! \! \left[ q_{\mu\nu}(x) +
q_{\mu\nu}(x-e_\nu) \right]^2 \ \right] \ . \nonumber
\end{eqnarray}
As can be read off from the first expression, for each link in the lattice,
all pairs of elementary squares attached to that link, but not lying in
one plane, are examined. If both members of such a pair are part
of a lattice surface, this costs an action increment $c$. Thus, vortex
surfaces are penalized for ``turning a corner'', i.e., for their
curvature.

It should be noted that also an action term of the Nambu-Goto type,
proportional to the world-surface area, was considered for the
$SU(2)$ and $SU(3)$ models \cite{m1,su3conf}. The result of these
considerations is that, in practice, the effect of such a term can be
absorbed into the curvature term. Including it does not enhance the
phenomenological flexibility of the models appreciably, and it was
therefore ultimately dropped. In particular, no indication has been
found that a world-surface area term would be useful for the purpose
of driving the deconfinement phase transition towards first-order
behavior, as is necessary for an accurate modeling of $Sp(2)$
Yang-Mills theory.

The desired first-order behavior therefore has to be generated by
different dynamics. A promising strategy in this regard is suggested
by the experience gathered with the $SU(4)$ random vortex world-surface
model \cite{su4}. Also in that case, it was necessary to devise dynamics
which enhance the first-order character of the deconfinement transition.
A viable solution was found to be an action term which enhances vortex
branching. $SU(4)$ Yang-Mills theory allows for two physically distinct
types of center vortices, and the associated chromomagnetic flux can
combine and disassociate, thus creating a branched structure of the
world-surface configurations. In the present case, there is only one
type of vortex flux, and branching is consequently impossible. However,
an effect reminiscent of branching behavior can be envisaged: In terms
of world-surfaces composed of elementary squares on a lattice, branching
essentially means that more than two squares attached to a given link
are part of a vortex. This notion can indeed be translated to the case
studied here, albeit with a different physical interpretation. In the
present vortex model, it is possible for two (or even three) vortex
surfaces to share a lattice link; in this case, four (or even six)
elementary squares attached to the link are part of a vortex.
In terms of the propagation of vortex lines in three dimensions, this
corresponds to two (or even three) lines meeting at a point in
three-dimensional space and remaining attached to one another for a
finite length in the fourth direction before separating again. Enhancing
such behavior can be interpreted as making the vortices more ``sticky''.
Therefore, a promising avenue is the introduction of a vortex stickiness
term into the action,
\begin{equation}
S_s [q] = \sum_{x} \sum_{\mu } F \left( \sum_{\nu \neq \mu }
(q_{\mu \nu } (x) + q_{\mu \nu } (x-e_{\nu } ) ) \right)
\label{sstick}
\end{equation}
where
\begin{equation}
F(4)=s_4 \ , \ \ \ \ F(6)=s_6 \ , \ \ \ \ F=0 \ \ \mbox{else} \, .
\end{equation}
Thus, for each link in the lattice, if four elementary squares attached to
the link are part of a vortex, this is weighted by an action increment
$s_4 $; if six elementary squares attached to the link are part of a
vortex, this is weighted by an action increment $s_6 $. Choosing negative
values of $s_4 $ and $s_6 $ facilitates such behavior, corresponding to the
vortices becoming more sticky. In general, $s_4 $ and $s_6 $ are
independent parameters. However, for the remainder of the present
investigation, only the case $s_4 \equiv s_6 \equiv s$ is considered
further, with the complete action
\begin{equation}
S[q] = S_c [q] + S_s [q]
\end{equation}
depending on two dimensionless parameters $c$ and $s$. In the absence
of the term $S_s [q]$, the deconfinement phase transition is second order,
and a viable model for the infrared sector of $SU(2)$ Yang-Mills theory
is achieved \cite{m1} for $c=0.24$. As will be seen below, introducing
$S_s [q]$ indeed serves to induce first-order behavior at the deconfinement
transition for sufficiently negative $s$, and the characteristics of
$Sp(2)$ lattice Yang-Mills theory listed in section \ref{sp2data} can be
reproduced.

\begin{figure}[h]
\centerline{\epsfig{file=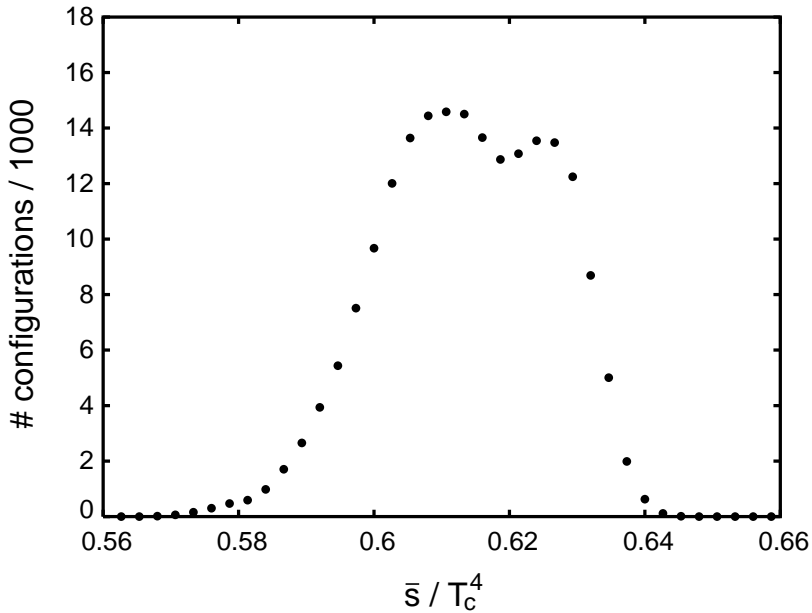,width=11cm} }
\caption{Distribution of the four-dimensional action density $\bar{s} $
at the deconfinement phase transition, for coupling parameters
$c=0.3394$ and $s=-1.24$. The measurement was taken on a
$50^3 \times 1$ lattice.}
\label{f1}
\end{figure}

\begin{figure}[t]
\centerline{\epsfig{file=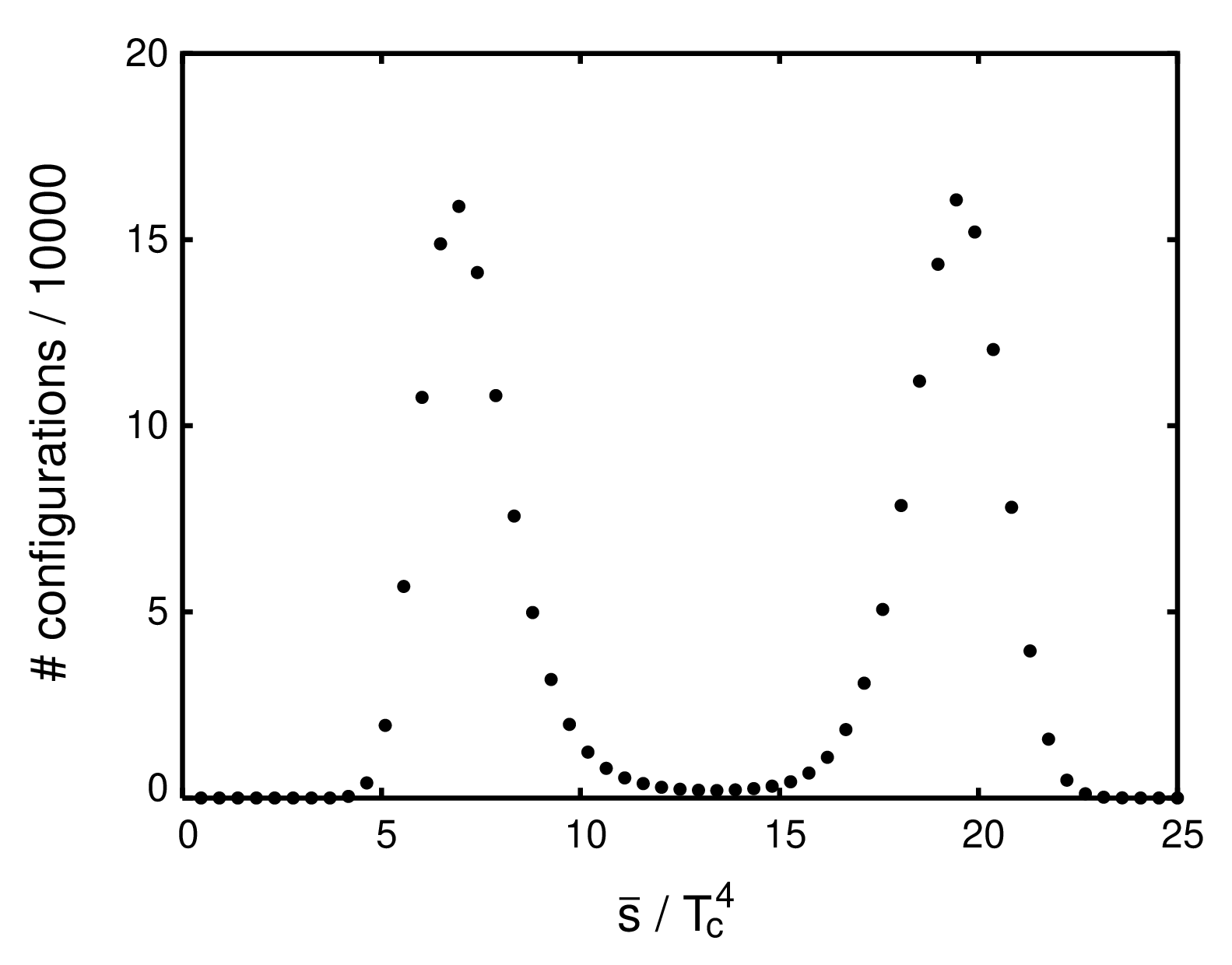,width=11cm} }
\caption{Distribution of the four-dimensional action density $\bar{s} $
at the deconfinement phase transition, for coupling parameters
$c=0.5469$ and $s=-1.99$. The measurement was taken on a
$6^3 \times 2$ lattice.}
\label{f2}
\end{figure}

\begin{figure}[t]
\centerline{\epsfig{file=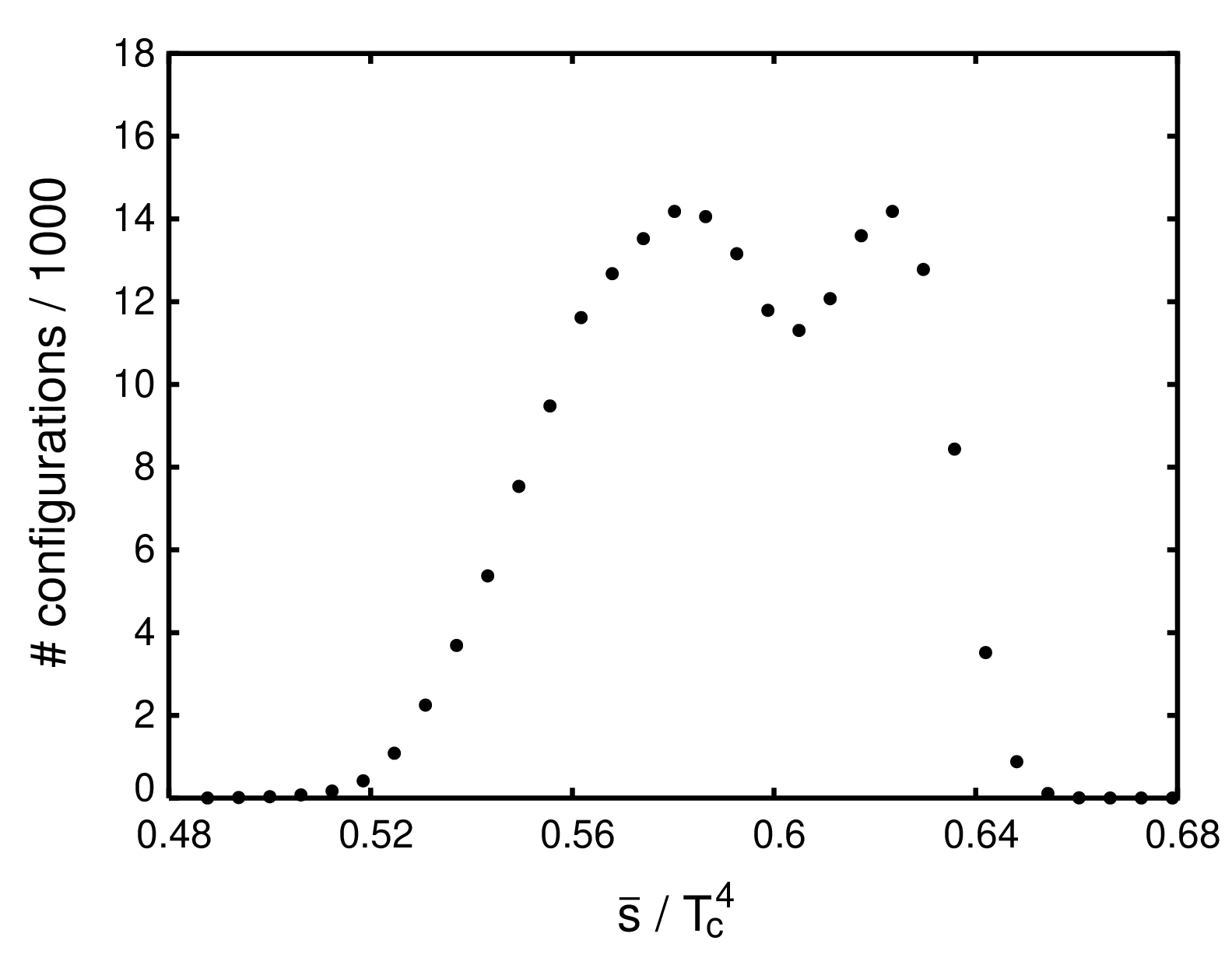,width=11cm} }
\caption{Distribution of the four-dimensional action density $\bar{s} $
at the deconfinement phase transition, for coupling parameters
$c=0.3513$ and $s=-1.3$. The measurement was taken on a
$30^3 \times 1$ lattice.}
\label{f3}
\end{figure}

\begin{figure}[t]
\centerline{\epsfig{file=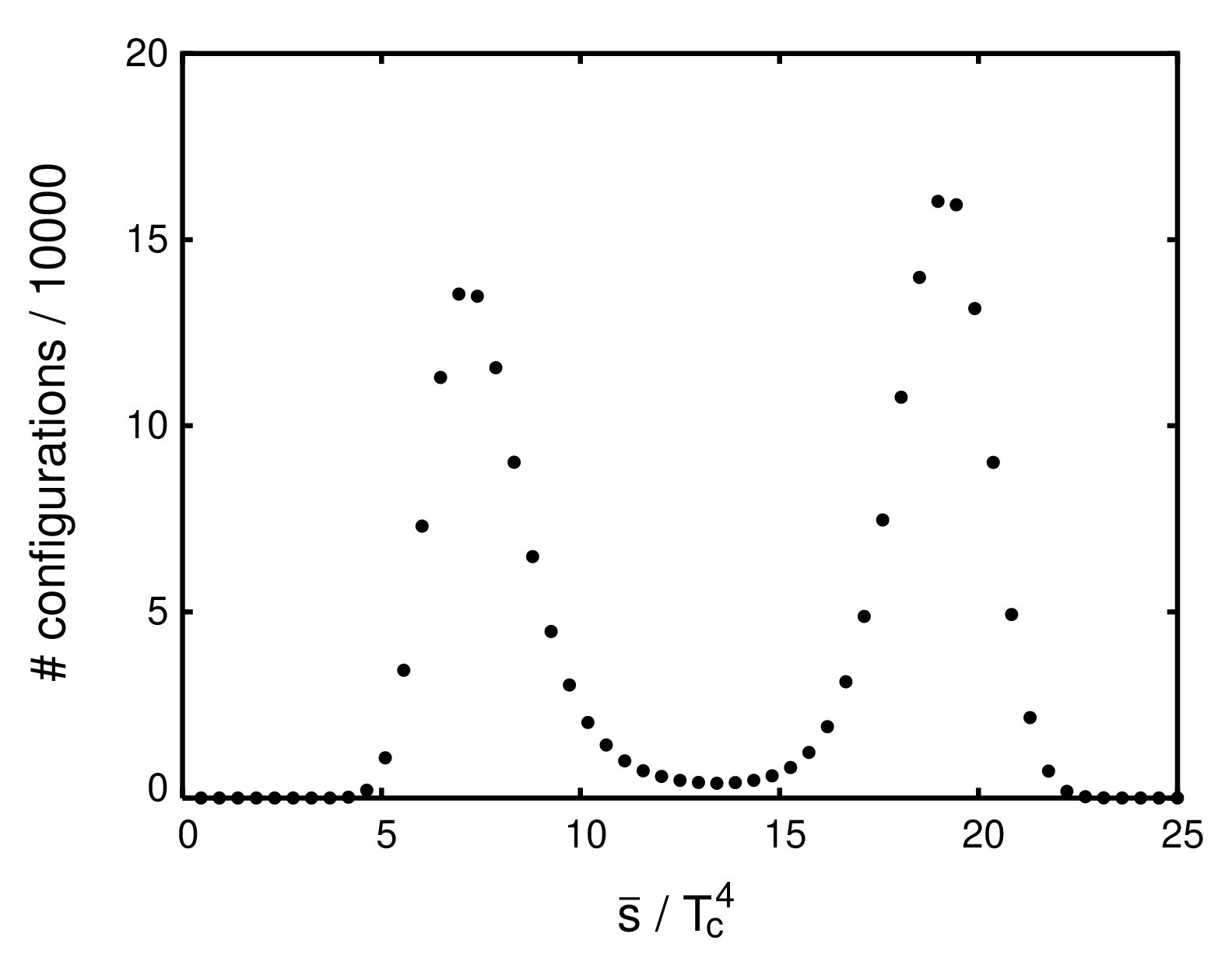,width=11cm} }
\caption{Distribution of the four-dimensional action density $\bar{s} $
at the deconfinement phase transition, for coupling parameters
$c=0.5337$ and $s=-1.9$. The measurement was taken on a
$6^3 \times 2$ lattice.}
\label{f4}
\end{figure}

\begin{figure}[t]
\centerline{\epsfig{file=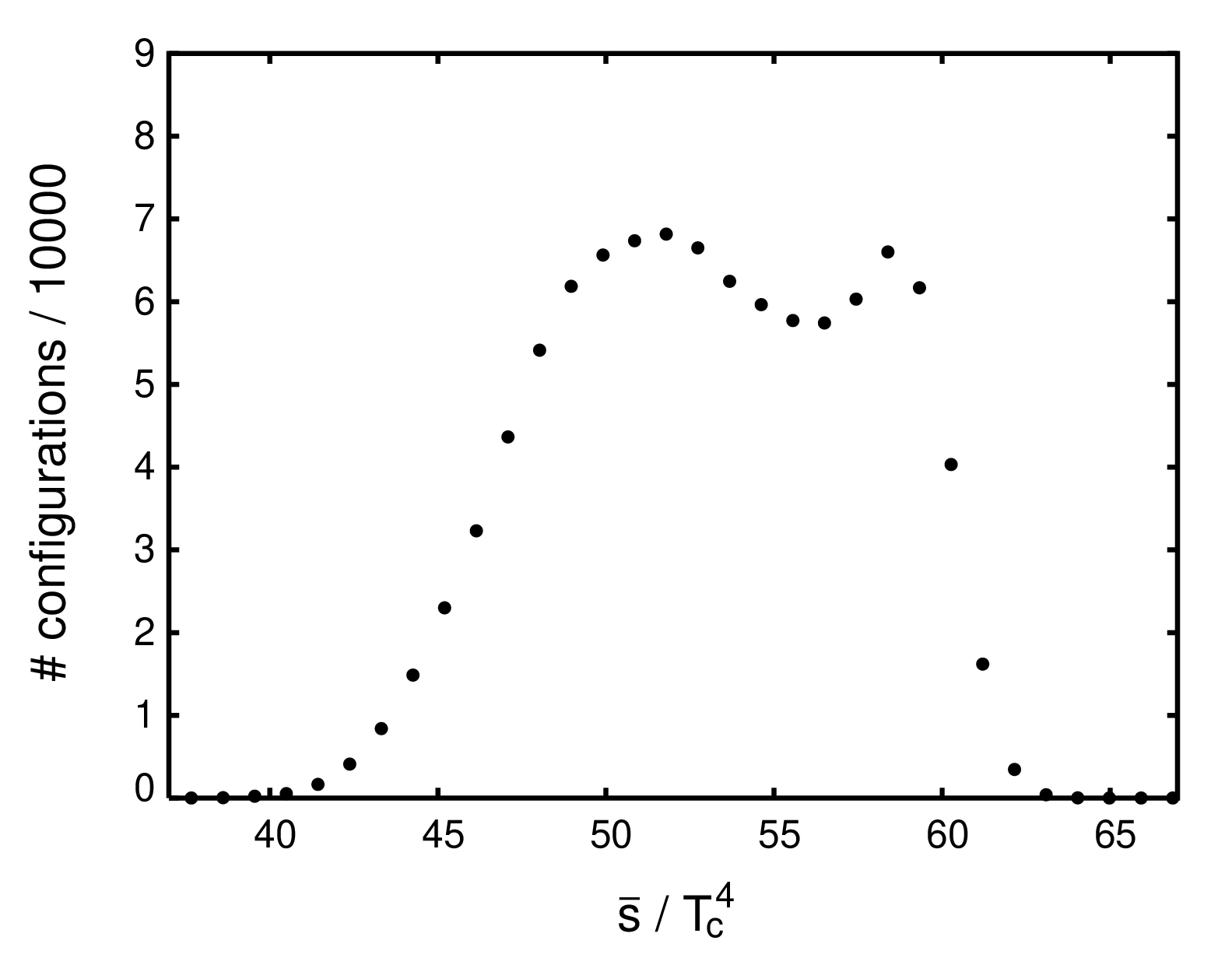,width=11cm} }
\caption{Distribution of the four-dimensional action density $\bar{s} $
at the deconfinement phase transition, for coupling parameters
$c=0.4546$ and $s=-0.3$. The measurement was taken on a
$16^3 \times 3$ lattice.}
\label{f5}
\end{figure}

\section{Locating the physical point}
\label{seclocat}
On the basis of the above model definition, Monte Carlo measurements of
observables can be carried out. The observables relevant for the
comparison with the $Sp(2)$ lattice Yang-Mills data given in section
\ref{sp2data} are, on the one hand, the probability distribution of the
action density and, on the other hand, Wilson loops, from which string
tensions can be extracted. The value of a Wilson loop in any given
vortex configuration is determined by the defining property of vortex
flux: Each instance of a vortex surface piercing an area spanned by
the loop\footnote{Wilson loops are defined on a lattice which is dual
to the one on which the vortices are constructed; thus, vortex piercings
of areas spanned by Wilson loops are defined unambiguously.} contributes a
phase factor $(-1)$ to the value of the Wilson loop. The action density
probability distribution is used to detect a first-order deconfinement
phase transition via a double-peak structure signaling the coexistence
of two phases. Examples of such action density distributions are given
in Figs.~\ref{f1}-\ref{f5}. The distance between the peaks in these
distributions gives a measure for the action density discontinuity
$\Delta \bar{s} $ at the transition. The corresponding values extracted
from Figs.~\ref{f1}-\ref{f5} are reported in Tables \ref{physpt2} and
\ref{physpt3} further below.

Since the lattice spacing in random vortex world-surface models is a
fixed physical quantity, only a discrete set of temperatures can be
accessed for a given set of coupling parameters $c$ and $s$. Therefore,
in general one cannot expect to realize the deconfinement transition
directly at the physical values of $c$ and $s$ which correctly model
full $Sp(2)$ lattice Yang-Mills theory; the inverse deconfinement
temperature usually will not be an integer multiple of the lattice
spacing at the physical point. For this reason, one has to resort to
an interpolation procedure \cite{m1}: The deconfinement transition is
studied at unphysical values of $c$ and $s$, on lattices extending a
varying number $N_t $ of spacings in the (Euclidean) time direction,
and the properties of the transition at the physical point are obtained
by interpolation. In the following, two such schemes will be investigated,
one based on lattices with $N_t =1,2$ and one based on lattices with
$N_t =1,2,3$. By trial and error, one can find sets of coupling parameters
$c$ and $s$ which yield a double peak in the action density distribution,
i.e., which realize the deconfinement transition, cf.~Figs.~\ref{f1}-\ref{f5}.
For these parameter sets, one therefore knows the deconfinement
temperature $T_c $ in lattice units, $aT_c =1/N_t $, and one can read
off the action density discontinuity in lattice units $a^4 \Delta \bar{s} $,
in which $a$ can be eliminated in favor of $T_c $. Measuring in addition
the zero-temperature string tension in lattice units, $\sigma a^2 $, one
can furthermore determine the ratio $T_c /\sqrt{\sigma } $. Such data
sets are given below in Tables \ref{physpt2} and \ref{physpt3}.

Before discussing these data sets, it should be noted that the requirement
of finding the deconfinement phase transition on a given lattice fixes only
one of the two coupling parameters $c$ and $s$. For a wide range of $c$,
one can find an appropriate $s$ realizing the transition, and vice versa.
The pairs of $c$ and $s$ for which data are reported below were singled out,
through extensive trial and error, by the additional requirement that
interpolation of these data sets must indeed yield the physical point,
i.e., must simultaneously yield the correct $Sp(2)$ values for
$T_c /\sqrt{\sigma } $ and $\Delta \bar{s} /T_c^4 $ given in section
\ref{sp2data}. Choosing a suitable point on an interpolation trajectory
always allows one to fit one of those values, but there is no guarantee
that the other one will simultaneously be correct. This is a nontrivial
additional constraint requiring a substantial search in the space of
coupling parameters $c$ and $s$. The final result of that search is the
specific set of data reported in Tables \ref{physpt2} and \ref{physpt3}.

\begin{table}[hb]
\begin{center}
\begin{tabular}{|c|c|c|c|c|}
\hline
$c$ & $s$ & $aT_c $ & $\Delta \bar{s} / T_c^4 $ & $T_c /\sqrt{\sigma } $ \\
\hline \hline
0.3394 & -1.24 & 1 & 0.014 & 0.816 \\
\hline
0.5469 & -1.99 & 0.5 & 13 & 0.474 \\
\hline
\end{tabular}
\end{center}
\caption{Sets of coupling parameters $c$, $s$ realizing the deconfinement
phase transition on lattices with $N_t =1,2$, together with values for
the action density discontinuity extracted from Figs.~\ref{f1} and \ref{f2}
and measurements of the ratio of the deconfinement temperature to the
square root of the zero-temperature string tension.}
\label{physpt2}
\end{table}

Table \ref{physpt2} displays suitable data sets found on lattices with
$N_t =1,2$. Since two data points are available for each quantity, all
quantities can be interpolated as linear functions of one parameter.
Choosing that parameter to be one of the relevant observables,
$T_c /\sqrt{\sigma } $, one immediately verifies that
$\Delta \bar{s} /T_c^4 = 8.76$ is indeed realized for
$T_c /\sqrt{\sigma } \approx 0.59$, as required by (\ref{input1}) and
(\ref{input2}). Similarly, for that value of $T_c /\sqrt{\sigma } $,
the coupling parameters $c$ and $s$ interpolate to
\begin{equation}
c=0.479 \ , \ \ \ \ \ \ \ \ \ \ \ \ s=-1.745 \ ,
\label{physparms}
\end{equation}
defining their physical values. Finally, $aT_c $ as a linear function of
$T_c /\sqrt{\sigma } $ interpolates to
\begin{equation}
aT_c = 0.663 \ ,
\label{physatc}
\end{equation}
implying that the inverse deconfinement temperature lies between $a$ and
$2a$, but is near neither of those two values; the physical point is not
close to either of the two data sets listed in Table \ref{physpt2}. This
is different from the $SU(N)$ models investigated in \cite{m1,su3conf,su4};
in those cases, the physical point is very near the $N_t =2$ data set and
the interpolation procedure only introduces small corrections to that set
in defining the physical point. In the present $Sp(2)$ case, the
uncertainty inherent in the interpolation procedure is much more
substantial due to the distance of the physical point from any of
the data sets reported in Table \ref{physpt2}.

One straightforward consistency check of the interpolation can be made
as follows. Up to this point, all quantities at the physical point have
been defined by interpolation of the data in Table \ref{physpt2}.
In the case of the deconfinement transition characteristics, one has
no choice in the matter, since these are not directly accessible at the
physical point (\ref{physparms}). However, the zero-temperature string
tension can be measured independently directly at the physical point.
The result of such a measurement, combined with (\ref{physatc}), again
yields the correct value $T_c /\sqrt{\sigma } =0.59$, buttressing the
validity of the interpolation procedure.

\begin{table}[ht]
\begin{center}
\begin{tabular}{|c||c|c|c|c|}
\hline
$N_t $ & $c$ & $s$ & $\Delta \bar{s} / T_c^4 $ & $T_c /\sqrt{\sigma } $ \\
\hline \hline
1 & 0.3513 & -1.3 & 0.040 & 0.810 \\
\hline
2 & 0.5337 & -1.9 & 12 & 0.485 \\
\hline
3 & 0.4546 & -0.3 & 7.3 & 0.480 \\
\hline
\end{tabular}
\end{center}
\caption{Sets of coupling parameters $c$, $s$ realizing the deconfinement
phase transition on lattices with $N_t =1,2,3$, together with values for
the action density discontinuity extracted from Figs.~\ref{f3}-\ref{f5}
and measurements of the ratio of the deconfinement temperature to the
square root of the zero-temperature string tension.}
\label{physpt3}
\end{table}

On the other hand, another way to gain insight into the uncertainty of
the interpolation lies in using an expanded data set obtained on lattices
with $N_t =1,2,3$ and comparing with the results obtained above.
Table \ref{physpt3} displays corresponding suitable data. Since three
data points are available for each quantity, all quantities can be
interpolated as parabolas depending on one parameter. Here, a difficulty
arises which is not present in the linear interpolation scheme discussed
further above: One cannot simply choose either of the two physical
dimensionless ratios $T_c /\sqrt{\sigma } $ or $\Delta \bar{s} /T_c^4 $
as the interpolation parameter, because the former is very closely spaced
between the $N_t =2$ and the $N_t =3$ data sets, leading to an extremely
unstable interpolation, and the latter is not even monotonous as $N_t $
rises. Consequently, to have well-spaced interpolation points conducive to
a stable interpolation, in the present case, $N_t $ was used as the
interpolation parameter. The drawback is, of course, that the entire
procedure becomes more indirect; both of the quantities of primary
interest, $T_c /\sqrt{\sigma } $ and $\Delta \bar{s} /T_c^4 $, are
interpolated as a function of a third parameter, instead of one being
interpolated directly as a function of the other. Constructing the
corresponding parabolas in $N_t $, one indeed verifies that the
relations (\ref{input1}) and (\ref{input2}) for $T_c /\sqrt{\sigma } $
and $\Delta \bar{s} /T_c^4 $ are simultaneously satisfied for
$N_t = 1.5573$. Furthermore, at that value of $N_t $, the parabolas for
the coupling parameters $c$ and $s$ yield
\begin{equation}
c=0.485 \ , \ \ \ \ \ \ \ \ \ \ \ \ s=-1.90 \ ,
\label{altparms}
\end{equation}
defining the physical point in this interpolation scheme. Finally,
identifying $aT_c =1/N_t $, one has at the physical point
\begin{equation}
aT_c = 0.642 \ .
\label{altatc}
\end{equation}
Also this interpolation scheme can be cross-checked by independently
calculating the zero-temperature string tension in lattice units,
$\sigma a^2 $, at the point (\ref{altparms}) in the space of coupling
parameters, and combining this with (\ref{altatc}) to obtain another
determination of $T_c /\sqrt{\sigma } $. This yields the value
$T_c /\sqrt{\sigma } = 0.52$, deviating significantly from the
interpolated value $T_c /\sqrt{\sigma } = 0.59$. Despite extensive search
in the space of coupling constants, the authors were unable to find a more
consistent data set. Thus, the interpolation scheme using $N_t =1,2,3$
appears to be less reliable\footnote{In general, there is no guarantee
that the interpolating polynomial becomes more accurate as more values of
$N_t $ are added to the data set, especially values which are distant from
the physical point; interpolation can become less stable as its order is
increased. Also from a more physical point of view, at $N_t =3$, the
vortex model constructed here is already rather far removed from the
infrared $Sp(2)$ physics of interest, and including data from this case
may well have the effect of distorting the physical picture rather than
improving convergence.} than the one using $N_t =1,2$. In view of this,
the set of coupling constants (\ref{physparms}) will be regarded in the
following as the best approximation to the physical point, and deviations
obtained using the set (\ref{altparms}) will be taken as an indication of
the systematic uncertainty inherent in the interpolation procedure.
Comparing (\ref{physparms}) with (\ref{altparms}), as well as
(\ref{physatc}) with (\ref{altatc}), these uncertainties appear to
be under 10\%. A further such comparison will be possible for the
spatial string tensions discussed below, similarly leading to an error
estimate of around 10\%.

\section{Predictions for the spatial string tension}
On the basis of the model for the infrared sector of $Sp(2)$ Yang-Mills
theory constructed above, predictions of further physical quantities
can be made. One important nonperturbative characteristic of Yang-Mills
theory is the behavior of the spatial string tension $\sigma_{S} $ at
finite temperatures. Using the physical set of coupling parameters
(\ref{physparms}), measurements on lattices with $N_t =1,2,3$ yield the
results listed in Table \ref{spats}, where $N_t $ has been translated
into $T/T_c $ using (\ref{physatc}).

\begin{table}[h]
\begin{center}
\begin{tabular}{|c||c|c|c|}
\hline
$T/T_c $ & 0.50 & 0.75 & 1.51 \\
\hline \hline
$\sigma_{S} (T) / \sigma_{S} (T=0)$ & 1.00 & 1.02 & 1.36 \\
\hline
\end{tabular}
\end{center}
\caption{Predictions for the behavior of the spatial string tension
$\sigma_{S} $ at finite temperatures, normalized to the zero-temperature
value $\sigma_{S} (T=0) \equiv \sigma $.}
\label{spats}
\end{table}

The characteristic rise of the spatial string tension in the deconfined
phase observed in $SU(N)$ Yang-Mills theories is predicted to occur also
in the $Sp(2)$ case, Table \ref{spats} giving a quantitative measure
for this behavior. By carrying out corresponding measurements within
$Sp(2)$ lattice Yang-Mills theory, the validity of the vortex model
constructed here can be put to test. To obtain an indication of the
systematic uncertainty in the above predictions, engendered by the
interpolation procedure used in defining the physical point, it is
useful to calculate the spatial string tension also for the alternate
set of coupling parameters (\ref{altparms}). This yields the results
displayed in Table \ref{altspats}.

\begin{table}[h]
\begin{center}
\begin{tabular}{|c||c|c|c|}
\hline
$T/T_c $ & 0.52 & 0.78 & 1.56 \\
\hline \hline
$\sigma_{S} (T) / \sigma_{S} (T=0)$ & 1.00 & 1.01 & 1.2 \\
\hline
\end{tabular}
\end{center}
\caption{Behavior of the spatial string tension $\sigma_{S} $ at finite
temperatures, normalized to the measured zero-temperature value
$\sigma_{S} (T=0) \equiv \sigma $, for the alternate set of coupling
parameters (\ref{altparms}). Deviations compared to Table \ref{spats}
give an indication of the systematic uncertainty in predicting the
spatial string tension.}
\label{altspats}
\end{table}

When using the coupling parameters (\ref{altparms}), the discussion
following eq.~(\ref{altatc}) should be kept in mind: Already the
measurement of the zero-temperature string tension using (\ref{altparms})
leads, combined with (\ref{altatc}), to a significant deviation from
the correct value $T_c /\sqrt{\sigma } = 0.59$. Thus, the spatial string
tension measurement at finite temperatures can be expected to suffer
from similar distortions. Table \ref{altspats} therefore gives the ratio
of the finite-temperature spatial string tension measured using
(\ref{altparms}) to the zero-temperature string tension {\em measured}
using (\ref{altparms}). This should cancel the distortions to some extent;
in particular, it leads to the correct low-temperature limit. For the
highest temperature displayed, Table \ref{altspats} displays a ratio
which is roughly 10\% below the value in Table \ref{spats}. In
comparison, if one used a value for the zero-temperature string tension
consistent with $T_c /\sqrt{\sigma } = 0.59$, then that ratio would rise
to a value roughly 10\% above the value in Table \ref{spats}. Altogether,
therefore, the systematic uncertainty also in the case of the predictions
given in Table \ref{spats} is of the order of 10\%, similar to the quantities
considered in section \ref{seclocat}.

\section{Conclusions}
The main objective of the present work was to demonstrate, by explicit
construction of a corresponding random vortex world-surface model, that
the center vortex picture can encompass the infrared physics of both
$SU(2)$ and $Sp(2)$ Yang-Mills theory. Doubts in this respect had recently
arisen in some quarters, based on the observation that the two Yang-Mills
theories contain the same center vortex degrees of freedom, and yet exhibit
qualitatively different behavior at the deconfinement phase transition, as
demonstrated in \cite{spn,g21o,pepelat05}. To resolve this apparent
dichotomy, it is necessary to take into account that, while $SU(2)$ and
$Sp(2)$ Yang-Mills theory contain the same center vortex degrees of freedom,
the effective actions governing those degrees of freedom are different;
after all, different cosets would have to be integrated out if one were
to derive those effective actions from the underlying Yang-Mills theories.
Thus, there is no obstacle in principle to both theories being described
by vortex models in the infrared sector; the present investigation set out
to show that such descriptions can indeed be achieved in practice.

Within the random vortex world-surface model, the vortex effective action
is determined phenomenologically. As shown in the present work, the
introduction of a vortex ``stickiness'' provides a way to drive the
deconfinement phase transition towards first-order behavior, which is
necessary for a correct description of the transition in the $Sp(2)$ case.
By adjusting the stickiness and curvature coefficients in the vortex
effective action, agreement with known data from $Sp(2)$ lattice Yang-Mills
theory was achieved, subject to systematic uncertainties engendered by
the interpolation procedure which is necessary to define the deconfinement
transition properties at the physical point. While these uncertainties
remained small in the $SU(N)$ random vortex world-surface models studied
previously \cite{m1,m2,m3,su3conf,su3bary,su3freee,su4}, in the $Sp(2)$
case, they are sizeable, and are estimated to amount to roughly 10 \% for
the observables studied here. Subject to this caveat, the results of the
present modeling effort indeed support the notion that $Sp(2)$ Yang-Mills
theory can be described in terms of vortex dynamics in the infrared,
along with the $SU(2)$ case investigated in \cite{m1,m2,m3}. The
predictions for the spatial string tension at finite temperatures
presented in Table \ref{spats} above provide a further opportunity to
test this notion, through comparison with corresponding measurements
within $Sp(2)$ lattice Yang-Mills theory.

\section*{Acknowledgments}
This work was supported by the U.S.~DOE under grants DE-FG03-95ER40965
(M.E.) and DE-FG02-94ER40847 (B.S.).


\begin{thebibliography}{99}
\bibitem{hooft} G.~'t~Hooft, Nucl.~Phys. {\bf B138} (1978) 1.
\bibitem{aharonov} Y.~Aharonov, A.~Casher and S.~Yankielowicz,
Nucl.~Phys. {\bf B146} (1978) 256.
\bibitem{cornold} J.~M.~Cornwall, Nucl.~Phys. {\bf B157} (1979) 392.
\bibitem{mack} G.~Mack, Phys.~Rev.~Lett. {\bf 45} (1980) 1378.
\bibitem{olesen} H.~B.~Nielsen and P.~Olesen,
Nucl.~Phys. {\bf B160} (1979) 380.
\bibitem{jg1} L.~Del~Debbio, M.~Faber, J.~Greensite and
\v{S}.~Olejn{\'\i}k, Phys.~Rev.~D {\bf 55} (1997) 2298.
\bibitem{jg2} L.~Del~Debbio, M.~Faber, J.~Giedt, J.~Greensite and
\v{S}.~Olejn{\'\i}k, Phys.~Rev.~D {\bf 58} (1998) 094501.
\bibitem{tk1} T.~G.~Kov\'{a}cs and E.~T.~Tomboulis,
Phys.~Rev.~D {\bf 57} (1998) 4054.
\bibitem{df1} P.~de~Forcrand and M.~D'Elia,
Phys.~Rev.~Lett. {\bf 82} (1999) 4582.
\bibitem{int} M.~Engelhardt, K.~Langfeld, H.~Reinhardt and O.~Tennert,
Phys.~Lett. {\bf B431} (1998) 141.
\bibitem{per} M.~Engelhardt, K.~Langfeld, H.~Reinhardt and O.~Tennert,
Phys.~Rev.~D {\bf 61} (2000) 054504.
\bibitem{jg3} J.~Greensite, Prog. Part. Nucl. Phys. {\bf 51} (2003) 1.
\bibitem{m1} M.~Engelhardt and H.~Reinhardt,
Nucl.~Phys. {\bf B585} (2000) 591.
\bibitem{m2} M.~Engelhardt, Nucl.~Phys. {\bf B585} (2000) 614.
\bibitem{m3} M.~Engelhardt, Nucl.~Phys. {\bf B638} (2002) 81.
\bibitem{su3conf} M.~Engelhardt, M.~Quandt and H.~Reinhardt,
Nucl.~Phys. {\bf B685} (2004) 227.
\bibitem{su3bary} M.~Engelhardt, Phys.~Rev.~D {\bf 70} (2004) 074004.
\bibitem{su3freee} M.~Quandt, H.~Reinhardt and M.~Engelhardt,
Phys.~Rev.~D {\bf 71} (2005) 054026.
\bibitem{cw2} J.~M.~Cornwall, Phys.~Rev.~D {\bf 61} (2000) 085012.
\bibitem{contvort} M.~Engelhardt and H.~Reinhardt,
Nucl.~Phys. {\bf B567} (2000) 249.
\bibitem{rb} R.~Bertle, M.~Engelhardt and M.~Faber,
Phys.~Rev.~D {\bf 64} (2001) 074504.
\bibitem{bruck} F.~Bruckmann and M.~Engelhardt, Phys.~Rev.~D {\bf 68}
(2003) 105011.
\bibitem{su4} M.~Engelhardt, Phys.~Rev.~D~73, 034015 (2006).
\bibitem{jeffstef} J.~Greensite and \v{S}.~Olejn{\'\i}k,
JHEP {\bf 0209} (2002) 039.
\bibitem{spn} K.~Holland, M.~Pepe, and U.-J.~Wiese,
Nucl.~Phys. {\bf B694} (2004) 35.
\bibitem{g21o} M.~Pepe, Nucl.~Phys.~Proc.~Suppl. {\bf 141} (2005) 238.
\bibitem{pepelat05} M.~Pepe, PoS {\bf LAT2005} (2005) 017.
\end{thebibliography}
\end{document}